\documentclass[floatfix,10pt,superscriptaddress,twocolumn,showpacs,showkeys,aps,prb,final]{revtex4-1}

\usepackage{graphicx}
\usepackage{dcolumn}
\usepackage{subfigure}
\usepackage{bm}
\usepackage{ifpdf,braket}
\usepackage{epstopdf}

\ifpdf
\usepackage[pdftex,
        colorlinks=true,
        pdftitle={Variational principle applied to two-band superconductors},
        pdfauthor={Antonio R. de C. Romaguera},
        pdfsubject={Vortex theory},
        pdfkeywords={Departamento de Fisica, Universidade Federal Rural de Pernambuco, 52171-900 Recife, Pernambuco, Brazil},
        baseurl={http://www.cmt.ua.ac.be},
        pdftoolbar=true,
        pdfmenubar=false,
        pdfpagemode=UseOutlines,
        pdfstartview=FitH,
        bookmarksopen=false
        ]{hyperref}
\fi

\begin{document}

\title{Evidence for skyrmions in the high-temperature superconductors}
\author{Alfredo A. Vargas-Paredes}%
\affiliation{Departamento de F\'{\i}sica dos S\'{o}lidos,
Universidade Federal do Rio de Janeiro, 21941-972 Rio de Janeiro,
Brazil}
\author{Marco Cariglia}%
\affiliation{Departamento de F\'{\i}sica , Universidade Federal de
Ouro Preto, 35400-000 Ouro Preto Minas Gerais, Brazil}
\author{Mauro M. Doria}%
\affiliation{Departamento de F\'{\i}sica dos S\'{o}lidos, Universidade Federal do Rio de Janeiro, 21941-972 Rio de Janeiro, Brazil}%
\author{Edinardo I. B. Rodrigues}%
\affiliation{Departamento de F\'{\i}sica, Universidade Federal Rural de
Pernambuco, 52171-900 Recife, Pernambuco, Brazil}
\author{A. R. de C. Romaguera}%
\affiliation{Departamento de F\'{\i}sica, Universidade Federal Rural de
Pernambuco, 52171-900 Recife, Pernambuco, Brazil}
\email{mmd@if.ufrj.br}

\date{\today}

\begin{abstract}
We claim that the charge density wave recently found by resonant
soft x-ray scattering in layered copper oxides is the tetragonal
symmetry defined by the distance between neighbor Cu$^{3+}$ ions in
the CuO$_2$ layer that determines the critical temperature. We find
evidence that this tetragonal symmetry is a skyrmionic state, which
is responsible for an unusual magnetic order and charge flow in the
layers that leads to the breaking of the time reversal symmetry
below the pseudogap line. The core of the skyrmions form pockets of
local magnetic field piercing the superconducting layers in opposite
direction to the rest of the unit cell.
\end{abstract}


\maketitle

\section{Introduction}

Since nuclear magnetic resonance studies have
revealed~\cite{miller11} that damped spin correlations exist across
the entire range of doping for the copper oxides, and with
sufficient intensity to mediate superconductivity, new questions
have been raised for the coexistence of superconductivity,
magnetism~\cite{berg09,vojta09,daou10} and
charge~\cite{Ghiringhelli17082012}.

Superconductivity in the copper oxides exists under a dome shaped
curve of the so-called temperature versus doping phase diagram. In
the absence of doping the copper oxides are antiferromagnetic Mott
insulators based on Cu$^{++}$ (d$^9$) spins.  Upon hole or electron
doping by chemical substitution or oxygenation at out-of-layer sites
this antiferromagnetic state is rapidly destroyed and its associated
Neel transition temperature brought to zero. Above a certain doping
level superconductivity emerges, and the critical temperature
($T_c$) increases with doping up to a maximum value, associated to
the optimal doping. Beyond this maximum value further doping results
in a reduction of $T_c$. Thus according to the doping level with
respect to the maximum $T_c$, the superconducting state is called
underdoped, optimally doped, and overdoped, respectively.
The pseudogap state emerges at a temperature $T^*$, that is claimed
to be a phase transition line~\cite{he11}. In the underdoped regime
this temperature is above $T_c$ and decreases with increasing doping
level. At some doping  $T^*=T_c$, and beyond, one expects that the
pseudogap line enters the superconducting dome to finally reach a
quantum critical point at $T=0$~\cite{berg09,vojta09,daou10}. The
microscopic nature of the pseudogap remains controversial, but some
of its characteristics are being unveiled by recent experiments.
Time-reversal symmetry is spontaneously broken below the pseudogap
line, because left-circularly polarized photons give a different
photocurrent from right-circularly polarized photons
\cite{kaminski02,xia08}. The Fermi surface breaks apart in the
pseudogap phase, above the superconducting critical temperature
$T_c$, leading to the emergence of electron pockets for hole-doped
cuprates \cite{norman10}. This transformation of the Fermi surface
has been suggested to be a consequence of a new periodicity that
sets in the system \cite{taillefer09}.
During the early days of superconductivity Bloch and Landau
suggested that superconductivity was governed by a ground state with
spontaneous circulating currents. This idea was soon to be disproved
by the fair argument that circulating currents have an energetic
cost that increase the kinetic energy. In terms of an order
parameter approach mean the presence of a gradient term, $\vec
\nabla \psi$. There is no question that a constant order parameter
is the preferred ground state solution in this case. Interestingly
many experiments done in the high- temperature superconductors
indicate the presence of a corrugated superconducting
state~\cite{Ghiringhelli17082012}, called by some as PDW (pairing
density state)~\cite{berg09,vojta09,daou10}. Thus for an
inhomogeneous state it is possible to entertain the idea of
spontaneously circulating currents in the ground state. Nevertheless
the question remains how to prevent their decay into the homogenous
state, which has lower energy. Here we claim that topological
stability is the key for the stability of heterogeneous
superconducting state. Therefore we claim that skyrmions are behind
the stability of corrugated superconducting states and set
topological protectorates.

In this paper we conjecture that near to the optimal doping (maximum
$T_c$), and below the pseudogap line, there is a tetragonal skyrmion
state in the copper oxides. Skyrmions have been found in several
non-superconducting compounds, such as the antiferromagnet
La$_2$Cu$_{1-x}$Li$_x$O$_4$~\cite{marcelo11}, the helimagnet
MnSi~\cite{pfleiderer04} and the doped semiconductor
Fe$_{1-x}$Co$_x$Si~\cite{munzer10}. In this paper we describe
qualitatively our conjecture of a skyrmion state based on an order
parameter approach~\cite{doria10,alfredo13}.
\section{Tetragonal symmetry}
Recently G. Ghiringhelli et al.~\cite{Ghiringhelli17082012}  have
found signals of a charge-density-wave instability that competes
with superconductivity  using resonant soft x-ray scattering. This
two-dimensional charge-density-wave in the underdoped compound
YBa$_2$Cu$_3$O$_{6+x}$ has an incommensurate periodicity of nearly
$3.2a$, (atomic unit cell parameters $a=0.39$ nm and $c=1.17$ nm, as
shown in Table 1 of Ref.~\cite{Ghiringhelli17082012}-supplementary
material and in Ref.~\cite{roeser2008a}). This periodicity sets a
tetragonal lattice because it is found to exist in orthogonal
directions, namely, along and perpendicular to the so-called CuO
chains, which act as charge reservoirs for the CuO$_2$ layer in this
material. Interestingly they find that this structure holds both
above and below $T_c$ and has correlations that reach the size of
$\bar \lambda =(16\pm 2)a$  for the underdoped compound
YBa$_2$Cu$_3$O$_{6.6}$. The compound with $T_c=57$ can be regarded
as the optimally doped in case of depleted CuO chains. We point out
that this resonant elastic x-ray scattering (REXS) correlation
length of $\bar \lambda =6.24\pm 0.78$ nm, found by  G. Ghiringhelli
et al., is equal, within error range, to the length that defines the
tetragonal structure at optimal doping (maximum $T_c$), claimed by
R.P. Roeser et al.\cite{roeser2008a}. According to R.P. Roeser et
al. this tetragonal structure ``acts as a resonator for a coherent
phase transition from a particle gas to a superconducting state" at
$T_c$. The average distance between neighbor Cu$^{3+}$ ions in the
CuO$_2$ layer of the YBa$_2$Cu$_3$O$_{6.5}$ compound is $L=3.49$ nm.
Basically H.P. Roeser et al. claim that this length really sets a de
Broglie wavelength $\lambda \equiv 2L$ that resonates in the unit
cell, namely, defines $T_c$ through the free particle in a box
expression, $\pi kT_c = p^2/2m$, where $p=h/\lambda$ and $m$ is the
Cooper pair mass. The relation $T_c \propto 1/L^2$ holds for the
copper oxide superconductors~\cite{roeser2011}. Thus we claim here
that the charge density correlation length of G. Ghiringhelli et al.
is nothing but the de Broglie wavelength of Roeser et al., namely,
$\bar \lambda = \lambda$. Next we argue that this tetragonal
symmetry is indicative of a skyrmionic state.

\section{Breaking of time reversal symmetry}
Below the pseudogap line one must seek a tetragonal symmetric state
that breaks the time-reversal symmetry. Next we show that this
breaking is a straightforward consequence of the following
assumptions: (i) superconductivity arises in the layers and
evanesces away from them; (ii) there are spontaneous superficial
currents ($\boldsymbol{J_s}$) confined to the CuO$_2$ layers (or a
superficial magnetization, $\boldsymbol{M_s}= - c\, \hat x_3 \times
\boldsymbol{J_s}$, where $c$ is the speed of light), and (iii) in
between the layers there may exist a metallic state able to sustain
three-dimensional superconductivity.  The breaking of the time
reversal symmetry is a straightforward consequence of
$\boldsymbol{J_s}$ in the layers, which produce a magnetic field,
$\boldsymbol{h}$, in space, whose parallel component,
$\boldsymbol{h}_{\parallel} \equiv h_1\hat x_1 + h_2\hat x_2$ is
discontinuous across the CuO$_2$ layer, while the perpendicular one,
$h_3$, is continuous. According to Maxwell's equations for a layer
at $x_3=0$, $\hat x_3 \times
[\boldsymbol{h}\left(0^+\right)-\boldsymbol{h}\left(0^-\right)] =
4\pi \boldsymbol{J_s}/c$, and $ \hat x_3 \cdot [\boldsymbol{
h}\left(0^+\right)-\boldsymbol{h}\left(0^-\right)] = 0$. A time
reversal operation ultimately corresponds to the exchange of t into
-t, but in a static treatment, this operation is simply played by
$\boldsymbol{h}$ into $-\boldsymbol{h}$. Hence time reversal is no
longer a symmetry nor is the spatial reflection $\hat x_3$
$\rightarrow$ $-\hat x_3$, under a fixed $\boldsymbol{J_s}$, but
their product is still a symmetry. The presence of spontaneous
circulating currents in the superconducting ground state has been
considered long ago by Felix Bloch and Lev Landau during the early
days of superconductivity \cite{schmalian10}. Recently C. M. Varma
\cite{varma06} has proposed that microscopic orbital currents within
the CuO$_2$ atomic cell describe properties of the pseudogap. In the
present approach we consider a state with tetragonal symmetry,
namely, with spontaneous circulating currents in an area larger,
than that claimed by Varma, which is confined to the atomic cell
($a$). The present area is defined by the optimal doping length
($L$). We claim here that $\boldsymbol{h}$, created by
$\boldsymbol{J_s}$ is such that there are skyrmions, defined by
their topological charge, given by,
\begin{eqnarray}\label{skyrmion}
Q= \frac{1}{4\pi}\int_{x_3=0} \big (\frac{\partial \hat h}{\partial
x_1} \times \frac{\partial \hat h}{\partial x_2} \big)\cdot \hat h
\; d^2x,
\end{eqnarray}
for $\hat h = \vec h/\vert \vec h \vert$.  Skyrmions cannot decay
into other configurations because of this topological stability no
matter how close they are in energy to any other configuration.
Clearly the time reversal symmetry, ($\vec h \rightarrow -\vec h$),
is broken by the skyrmions. The skyrmions are chiral magnetic
excitations with a well defined core, where the rotation sense is in
the opposite direction of the rest, such that there is no net
current flowing out of the unit cell area. This core has the size of
nearly $3a$, thus close to the proposed periodicity of G.
Ghiringhelli et al.~\cite{Ghiringhelli17082012}. In the tetragonal
symmetric state the cells carry skyrmions with topological charge
$Q$. The preferred chirality of the skyrmions will rotate circularly
polarized light passing through the layers and lead to the dichroism
observed below the pseudogap line\cite{xia08}.
\section{Structure of the skyrmion lattice}
For the compound YBa$_2$Cu$_3$O$_{6.5}$ the ratio between the unit
cell length and the atomic unit cell height is $L/c=2.98$
nm~\cite{roeser2008a}. We interpret this ratio as a consequence of
the skyrmion lattice. To reach this conclusion consider, for
simplicity, the copper oxyde superconductor as made of a stack of
identical CuO$_2$ layers separated by a fixed distance $c$. Thus in
each layer there is an identical tetragonal lattice whose cells have
$Q$ skyrmions each. Hence there is an elaborate spatially magnetic
field arrangement in between the layers whose magnetic field energy
density, $\boldsymbol{h}^2/8\pi$, stored in space, must be
considered. We find that this magnetic energy integrated over the
unit cell is optimally reduced for some values of the unit cell
length $L$, that will be discussed in detail in a future
publication. Thus the present approach shows that the distance $c$
between layers plays a key role in defining $L$, and consequently
$T_c$, as stressed by some authors~\cite{innocenti10}. According to
our theory this minimum is reached at $L/c \sim 3$ and $4$ for $d$
and $s$ wave symmetries, respectively. Table~\ref{table1} displays
this ratio $L/c$ for some single layered compounds, according to
Roeser et al. ~\cite{roeser2008a,roeser2011}, to show that its value
is similar for different compounds. We suggest that this ratio truly
reflects some inner property of the tetragonal state, since it falls
within our theoretical calculations, which are a consequence of the
stored magnetic energy of the skyrmion lattice.

\begin{table}[t]
\caption{Single layer cuprate superconductors containing the atomic
unit cell height ($c$ nm), the maximum critical temperature ($T_c$
K), the average dopant distance ($L$ nm) obtained from $L\equiv
(26.3\, \mbox{nm}K^{1/2})/\sqrt{T_c}$ and the ratio $L/c$.}
\centering
\begin{tabular}{ccccc}
\hline
\hline
\begin{minipage}[b]{0.40\linewidth} Compound \end{minipage}  &\begin{minipage}[b]{0.10\linewidth} $T_c$ (K)  \end{minipage} & \begin{minipage}[b]{0.10\linewidth} $L$ (nm) \end{minipage} & \begin{minipage}[b]{0.10\linewidth} $c$ (nm) \end{minipage}& \begin{minipage}[b]{0.10\linewidth} $L/c$  \end{minipage}\\
\hline
\hline
\begin{minipage}[b]{0.40\linewidth} Bi$_2$(Sr$_{1.6 }$La$_{0.4}$)CuO$_{6+\delta}$  \end{minipage} &
\begin{minipage}[b]{0.10\linewidth}34 \end{minipage}& \begin{minipage}[b]{0.10\linewidth} 4.51 \end{minipage}& \begin{minipage}[b]{0.10\linewidth}1.220\end{minipage}&\begin{minipage}[b]{0.10\linewidth} 3.70 \end{minipage}\\
\hline
\begin{minipage}[b]{0.40\linewidth} Tl$_2$Ba$_2$CuO$_6$  \end{minipage} &\begin{minipage}[b]{0.10\linewidth} 80 \end{minipage}&\begin{minipage}[b]{0.10\linewidth} 2.94 \end{minipage}& \begin{minipage}[b]{0.10\linewidth}1.162 \end{minipage}&\begin{minipage}[b]{0.10\linewidth} 2.53\end{minipage}
\\
\hline
\begin{minipage}[b]{0.40\linewidth}  HgBa$_2$CuO$_{4+\delta}$  \end{minipage} &\begin{minipage}[b]{0.10\linewidth} 95\end{minipage} & \begin{minipage}[b]{0.10\linewidth}2.70 \end{minipage}& \begin{minipage}[b]{0.10\linewidth}0.950 \end{minipage}&\begin{minipage}[b]{0.10\linewidth} 2.84\end{minipage}
\\
\hline
\hline
\end{tabular}
\label{table1}
\end{table}

\begin{figure}[t]
\includegraphics [width=0.75\linewidth]{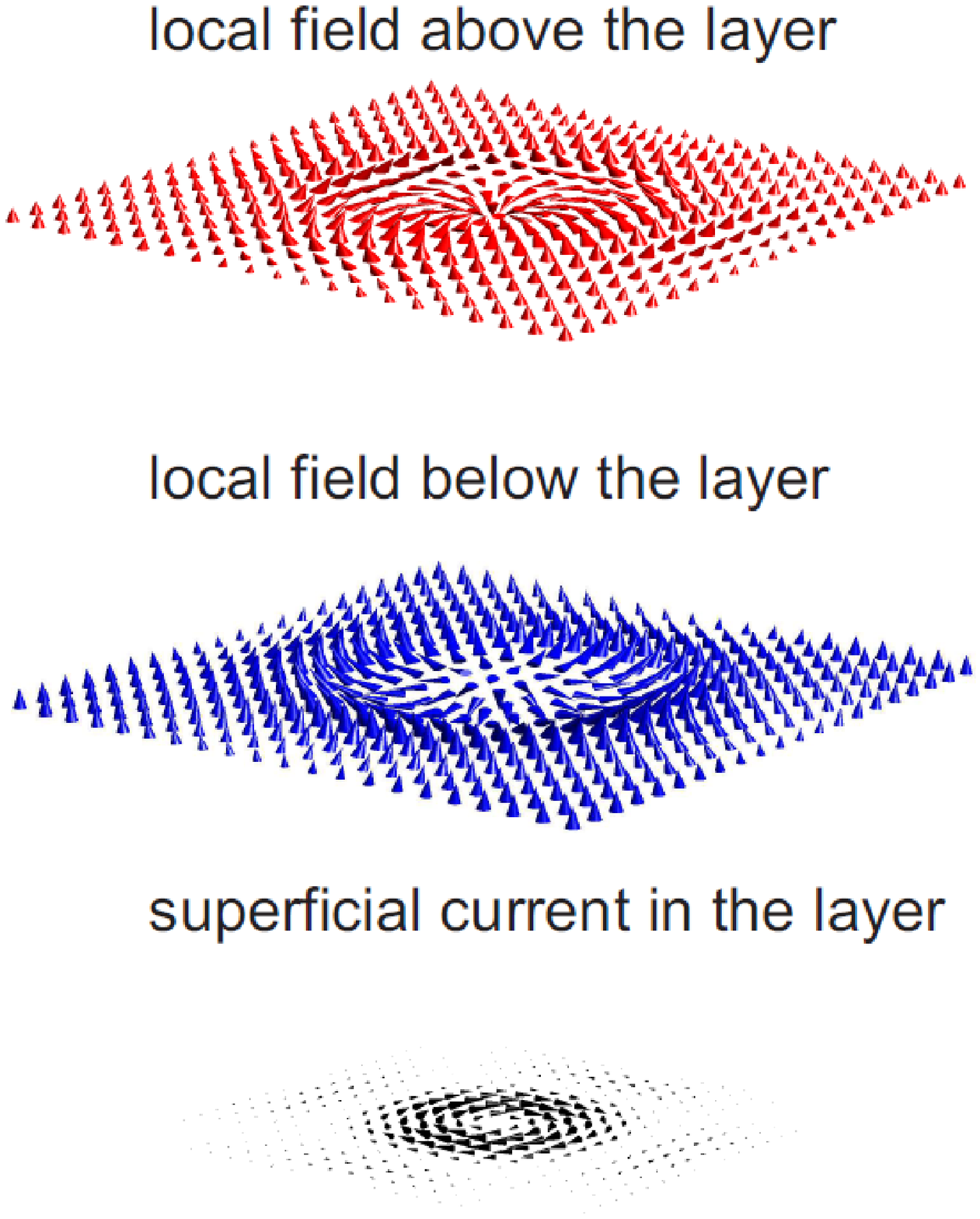}
\caption{ (color online) Pictorial view of the local magnetic field
of a skyrmion very near to a superconducting layer. Both views,
slightly above (red) and below (blue) the layer, are shown. The
discontinuity of the tangential field component results in the
superficial current which is also displayed here
(below).}\label{fig1}
\end{figure}

Let us analyze in more details the magnetic field associated to the
skyrmion assuming, for simplicity, that there should exist only two
kinds of magnetic field stream lines in the layered structure. Since
$\boldsymbol{\nabla \cdot h}=0$ there are those that pierce a
particular layer twice and those that pierce all the layers only
once, never to return again, like in a solenoid. Consider that the
skyrmion has a core, this core a center, and all the magnetic stream
lines of the first kind should pass through it. As shown in
Fig.~\ref{fig1} this center acts as a sinkhole, as seen from the top
of a layer, and a source, as seen from the bottom of the very same
layer. All the stream lines that pass through the core of the
skyrmion must return and cross the layer again piercing the
remaining area at some point in the opposite direction. Hence the
neighborhood of the skyrmion center, which is the core, has $h_3<0$,
while the remaining unit cell area has $h_3>0$. $\boldsymbol{J_s}$
around the center is very strong in comparison to the rest of the
unit cell, where it is weak and flows in the opposite direction,
thus the areas associated to flow and counterflow have very distinct
sizes. Thus the core of the skyrmion forms a pocket of local
magnetic field piercing the superconducting layers in opposite
direction to the rest. We determine that these pockets correspond to
9.4 $\%$ and 5.3 $\%$ of the total area, in case of d and s wave
respectively, for the optimal value of the unit cell size.

It must be emphasized that the local magnetic field created by the
skyrmion is very weak. Indeed it is fundamental to estimate the
value of $\boldsymbol{h}$ since NMR/NQR \cite{strassle08,strassle11}
and $\mu$SR \cite{macdougall08,sonier09} experiments put several
bounds on its maximum value. An estimate of the maximum modules of
$h_3$ can be obtained by calculating its value at the center of the
skyrmion. It follows by considering the core of the skyrmion as a
disk of radius $R$, such that $\boldsymbol{J_s}$ grows linearly
proportional away from the center of the skyrmion. The magnetic
moment generated by such a disk satisfies $\mu= h_3 \pi R^3/8$. This
relation shows that an extended object (assume, as an example, that
$R \sim L/4$, such that for $L\sim 3.0$ nm, $R\sim 0.75$ nm) can
have a large magnetic moment, $\mu \sim 0.01 \; \mu_B$, attached to
a very weak field, $h_3 \sim 0.02$ mT. This magnetic field and
magnetic moment is in the threshold limit of local fields from
dynamical currents measured by NMR/NQR \cite{strassle08,strassle11}
and by $\mu$SR \cite{macdougall08,sonier09} measurements. Because of
the elaborate circulating currents, the skyrmion lattice manifests
itself as a periodic charge arrangement.

\section{Conclusions}
We claim that there is a lattice of skyrmions in the Cu$O_2$ layers
that explains the tetragonal symmetry and the breaking of
time-reversal symmetry near to the optimal doping where the critical
temperature reaches a maximum.

\begin{acknowledgments}
Marco Cariglia was supported by Fapemig under project CEX APQ 2324-11. E. I. B. Rodrigues, A. R. de C. Romaguera, A. A. Vargas-Paredes and M. M. Doria acknowledge the Brazilian agencies CAPES, CNPq, FACEPE and FAPERJ for partial financial support.
\end{acknowledgments}

\bibliographystyle{ieeetr}
\bibliography{reference}

\end{document}